\documentclass[final,5p,twocolumn,times]{elsarticle}

\usepackage{amssymb}
\usepackage{graphicx}
\usepackage[latin1]{inputenc}
\usepackage{amsmath}
\usepackage{amsfonts}
\usepackage{amsbsy}
\usepackage{amssymb}
\usepackage{listings}
\usepackage{verbatim}
\usepackage{float}
\usepackage{bm}
\usepackage{color}
\usepackage{epstopdf}

\usepackage{relsize}
\usepackage{lipsum}

\def\ifmonospace{\ifdim\fontdimen3\font=0pt }

\def\C++{%
\ifmonospace%
    C++%
\else%
    C\kern-.1667em\raise.30ex\hbox{\smaller{++}}%
\fi%
\spacefactor1000 }
\newcommand{\NOTE}[1]{\textcolor{black}{#1}}

\newcounter{bla}

\journal{Computer Physics Communications}

\begin{document}

\begin{frontmatter}



\title{SoAx: A generic C++ Structure of Arrays for handling particles in HPC codes}


\author{Holger Homann\corref{author}}
\author{Francois Laenen}

\cortext[author] {Corresponding author.\\\textit{E-mail address:} holger.homann@oca.eu}
\address{Laboratoire J.-L.\ Lagrange, Universit\'e C\^ote d'Azur, Observatoire de la
     C\^ote d'Azur, CNRS, F-06304 Nice, France}

\begin{abstract}
  The numerical study of physical problems often require integrating
  the dynamics of a large number of particles evolving according to a
  given set of equations. Particles are characterized by the
  information they are carrying such as an identity, a position other.
  There are generally speaking two different possibilities for
  handling particles in high performance computing (HPC) codes. The
  concept of an \emph{Array of Structures} (AoS) is in the spirit of
  the object-oriented programming (OOP) paradigm in that the particle
  information is implemented as a structure. Here, an object
  (realization of the structure) represents one particle and a set of
  many particles is stored in an array. In contrast, using the concept
  of a \emph{Structure of Arrays} (SoA), a single structure holds
  several arrays each representing one property (such as the identity)
  of the whole set of particles. \\ The AoS approach is often
  implemented in HPC codes due to its handiness and flexibility. For a
  class of problems, however, it is know that the performance of SoA
  is much better than that of AoS. We confirm this observation for our
  particle problem. Using a benchmark we show that on modern Intel
  Xeon processors the SoA implementation is typically several times
  faster than the AoS one. On Intel's MIC co-processors the
  performance gap even attains a factor of ten. The same is true for
  GPU computing, using both computational and multi-purpose
  GPUs. \\ Combining performance and handiness, we present the library
  SoAx that has optimal performance (on CPUs, MICs, and GPUs) while
  providing the same handiness as AoS. For this, SoAx uses modern \C++
  design techniques such template meta programming that allows to
  automatically generate code for user defined heterogeneous data
  structures.

\end{abstract}

\begin{keyword}
  C++, Heterogeneous data, Template metaprogramming, Generic programming
\end{keyword}

\end{frontmatter}



{\bf PROGRAM SUMMARY/NEW VERSION PROGRAM SUMMARY}

\begin{small}
\noindent
{\em Manuscript Title:SoAx: A generic C++ Structure of Arrays for handling particles in HPC codes}                                       \\
{\em Authors: Holger Homann, Francois Laenen}                                                \\
{\em Program Title: SoAx}                                          \\
{\em Journal Reference:}                                      \\
{\em Catalogue identifier:}                                   \\
{\em Licensing provisions: CECILL}                                   \\
{\em Programming language:\C++}                                   \\
{\em Computer: CPU, MIC, GPU}                                      \\
{\em Operating system: Linux}                                 \\
{\em RAM: variable} bytes                                              \\
{\em Number of processors used: 1}                              \\
{\em Supplementary material:}                                 \\
{\em Keywords: C++, Heterogeneous data, Template metaprogramming, Generic programming}  \\
{\em Classification: 6.5 Software including Parallel Algorithms}                                         \\
{\em External routines/libraries:}                            \\
{\em Subprograms used:}                                       \\
{\em Catalogue identifier of previous version:}*              \\
{\em Journal reference of previous version:}*                  \\
{\em Does the new version supersede the previous version?:}*   \\
{\em Nature of problem: Structures of arrays (SoA) are generally faster
than arrays of structures (AoS) while AoS are more handy. This library
(SoAx) combines the advantages of both. By means of C++(11)
meta-template programming SoAx achieves maximal performance (efficient
use of vector units and cache of modern CPUs) while providing a very
convenient user interface (including object-oriented element handling)
and flexibility. It has been designed to handle list-like sets of
particles (similar to struct {int id; double[3] pos; float[3] vel;};)
in the context of high-performance numerical simulations. It can be
applied to many other problems.}\\
   \\
{\em Solution method: Template Metaprogramming, Expression Templates}\\
   \\
{\em Reasons for the new version:}*\\
   \\
{\em Summary of revisions:}*\\
   \\
{\em Restrictions:}\\
   \\
{\em Unusual features:}\\
   \\
{\em Additional comments:}\\
   \\
{\em Running time:}\\
   \\

* Items marked with an asterisk are only required for new versions
of programs previously published in the CPC Program Library.\\
\end{small}

\lstset{language=C++,frame=single}
\lstset{breaklines=true,basicstyle=\small}
\section{Introduction}

Particles are at the heart of many astrophysical, environmental or
industrial problems ranging from the dynamics of galaxies over
sandstorms to combustion in diesel engines. Investigating such
problems require generally integrating the dynamics of a large number
of particles evolving according to a given physical laws. Examples are
N-body simulations in cosmology \cite{bodenheimer2006numerical,
  teyssier:2012}, particle in cell codes (PIC) exploring plasma
physics \cite{grigoryev2002numerical, germaschewski-fox-etal:2016} or
hydrodynamic simulations studying Lagrangian turbulence problems
\cite{toschi-bodenschatz:2009, biferale:2004b}.  Such kind of
numerical simulations have in common that that they are numerically
expensive meaning that they rely on number crunching, i.e. an enormous
number of floating point operations. Studying the particle dynamics
during a finite time interval requires the numerical integration of
the underlying equations of motion over many time steps so that the
particle data (position, velocity, ...) is used in simple but numerous
repeated operations. The performance of such operations depend in a
crucial way on how the particle data is stored and accessed. \NOTE{For instance, the importance of spatial locality in memory access is a well known issue when it comes to the scalability of code performance on parallel architectures \citep{faria2013impact}}.

Modern supercomputers are often indispensable for studying challenging
problems. Their architecture got more and more complex in recent
years. The today's fastest computers consists of several performance
sensible components such as multi-level caches, vector units based on
the 'single-instruction multiple-data' (SIMD) concept, multi-core
processors, many-core (MIC) and GPU accelerators. Evidently it is
important to make use of all these components to optimize the
performance of a numerical code.

Particles can carry different properties such as an identity, a
position or a mass. In programming languages such as Fortran, C or
\C++, the data types \verb+int+, \verb+double+ and \verb+float+, could
be chosen to represent the former particle properties.  In this paper
codelets (serving as implementation examples) will always be given in
\C++, \NOTE{but the reasoning in section 1. and 2. will be kept general so
that it will similarly apply to Fortran and C}. 

In \C++, particles can be implemented as a heterogeneous structure
\begin{lstlisting}[caption={\label{lst:particle} Particle structure storing the data of one particle}]
  struct Particle {
    int id;
    double position;
    float mass;
  }
\end{lstlisting}
This way, individual particles can easily be generated as objects
(\verb+Particle p+;) and modified (\verb+p.id = 42+). A set of
particles is then often handled by an array- or list-like structures
(\verb+std::list<Particle> pList;+) providing functionalities such as
access, adding and removal of particles. Such an organization is
called \emph{array of structures} (AoS) as the particles are
represented by a structure that is hold by an array (or
list). Treating particles as objects is also convenient for
transferring them from one process to another via the message passing
interface (MPI) in parallel applications.

Another implementation strategy for handling a set of many particles
is to use one structure that holds several arrays; one array for each
particle property:
\begin{lstlisting}[caption={\label{lst:partArr} Structure of array containing one array per particle property}]
class PartArr 
{
public:
  int* id;
  double* position;
  float* mass;
}
\end{lstlisting}
It is then convenient to add member functions to this class that
perform operations on all the properties such as allocating memory:
\begin{lstlisting}[caption={\label{lst:partArr::alloc} Member function to allocate memory for particle property arrays}]
void PartArr::allocate(int n)
{
  number = new int[n];
  position = new double[n];
  mass = new float[n];
}
\end{lstlisting}
In the same way, member functions for adding, removing and other
functionalities could be added. This kind of implementation is called
\emph{structure of arrays} (SoA) from the fact that in this case one
structure handles a set of particles whose properties are represented
by different arrays. \verb+PartArr pArr;+ creates a set of particles
and an individual particle is referenced by the array index
(\verb+pArr.position[42]+ returns the position of particle 42). A
priory, particles cannot be extracted as individual objects from the
structure \verb+PartArr+. For this, a structure \verb+Particle+ (see
codelet above) would be needed together with a function copying the
array data for one index to the \verb+Particle+ member variables. From
these considerations it is clear that AoSs are easier to implement and
to use than SoAs.

\begin{figure}[h]
\centering
  \includegraphics[width=0.48\textwidth]{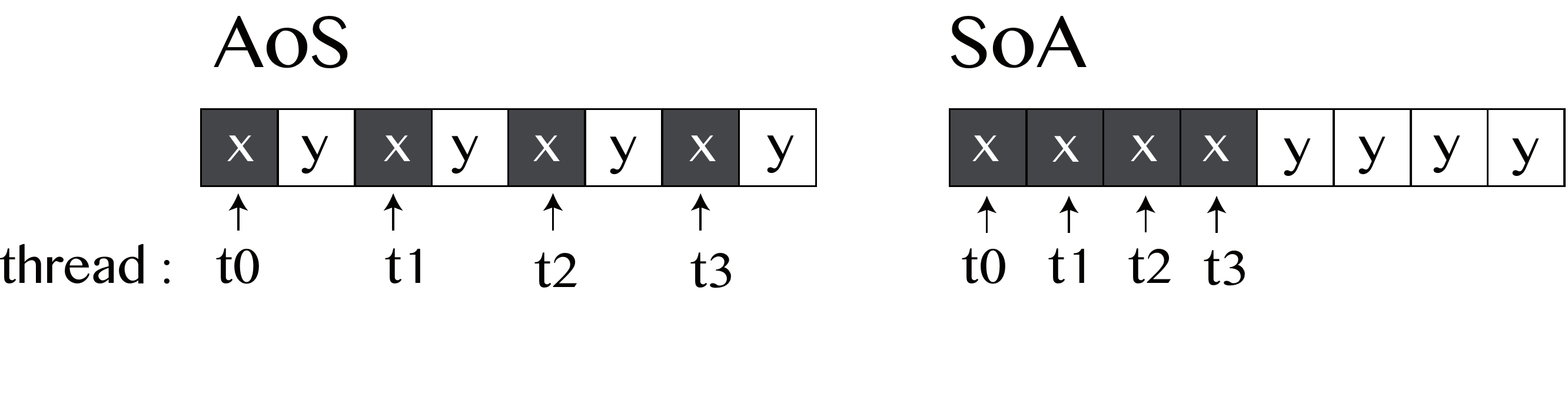}
  \caption{\NOTE{Illustration of the memory access by computing threads in the cases of AoS and SoA memory layouts. Usually, depending on the thread register size, more cache is wasted when accessing a data element (in this example, the x-coordinate of the particle).} }
\end{figure}

AoS are also more extendable than SoA. Imagine one would like to reuse
the above outlined particle storage implementation for a slightly
different particle type that requires the addition of a property such
as a charge. AoS are more flexible than SoA for this task: the novel
property could be added to the \verb+Particle+ structure by simply
adding the member variable \verb+float charge;+. In the case of a SoA
an array (through \verb+float* charge;+) could be added to
\verb+PartArr+. But in turn, all member functions such as
\verb+allocate+ would also have to be updated in order to treat the
added array.

AoS seem to be the better candidate to store particles than
SoA. However, SoA are faster in many circumstances (especially on MIC
and GPUs) \cite{faria-silva-etal:2013,
  huang-shi:2015,xue-yang-etal:2015} than AoS and we show that this is
also the case for typical manipulations (such as trajectory
integration) on particle data. By means of a benchmark modeling
floating-point operations used in real codes we show that SoAs are
typically several times faster than AoSs and that the performance of
an AoS depends on the size (in terms of bytes) of the structure
(\verb+Particle+ in the example above). In order to cope with the
seemingly contradicting properties handiness, flexibility and
performance, \NOTE{the implementation has to use advanced programming
  techniques. \C++ allows for abstractions that permit to access data
  via a AoS pattern while data is arranged in memory as a SoA
  \cite{Strzodka:2011}.} In this paper, we present another generic
implementation of a SoA called SoAx that has optimal performance while
providing the same handiness and flexibility of an AoS. 

This paper is organized as follows. In section \ref{sec:bench_CPU} we
benchmark the performance of AoS and SoA on CPUs. In section
\ref{sec:bench_GPU_MIC} we discuss a similar benchmark on GPUs and
MICs. The generic \C++ implementation of SoAx is presented in section
\ref{sec:soax}. Conclusions are drawn in \ref{sec:conclusion}.

\section{Benchmarking AoS and SoA on CPUs}
\label{sec:bench_CPU}

In order to compare the performance of SoA and AoS we measure the
execution time of a benchmark computation. The latter consists in
performing an Euler advection time step for the position $\bm{x}$ of a
set of particles
\begin{equation}
  \label{eq:euler}
  \bm{x} += dt\,\bm{v},
\end{equation}
$dt$ denoting the time step (a floating point number) and $\bm{v}$ the
velocity of the particle. This equation is a simple prototype for
typical operations appearing in numerical codes. It consists, for each
component, of two loads from the heap memory plus one for the constant
$dt$, usually from the stack, and one store in heap memory.

For benchmarking AoS we use the structure
\begin{lstlisting}[caption={\label{lst:particleBench} Particle structure used in the benchmark. SIZE is the number of supplementary floats.}]
template<int SIZE>
class Particle
{
  public:
  float x[3];
  float v[3];
  float temp[SIZE];
};
\end{lstlisting}
where temp is a place holder for additional particle properties that
might be necessary for the physical problem under consideration (such
as a mass, an electric charge...) or the numerical algorithm (such as
temporary positions and velocities for a Runge-Kutta scheme). In the
case of SoA we simply use three heap-allocated \C++ arrays for
$\bm{x}$ and $\bm{v}$, respectively.

Typically, in numerical simulations many successive time steps are
performed in order to integrate the particle dynamics. In our
benchmark we therefore loop many times over the numerical
implementation of (\ref{eq:euler}) \NOTE{and average. The code is
  compiled with gcc 5.1. As optimization flags we use -O3 and SIMD
  hardware specific flags such as -avx2.}

\begin{figure}[h]
  \centering
  \includegraphics[width=0.48\textwidth]{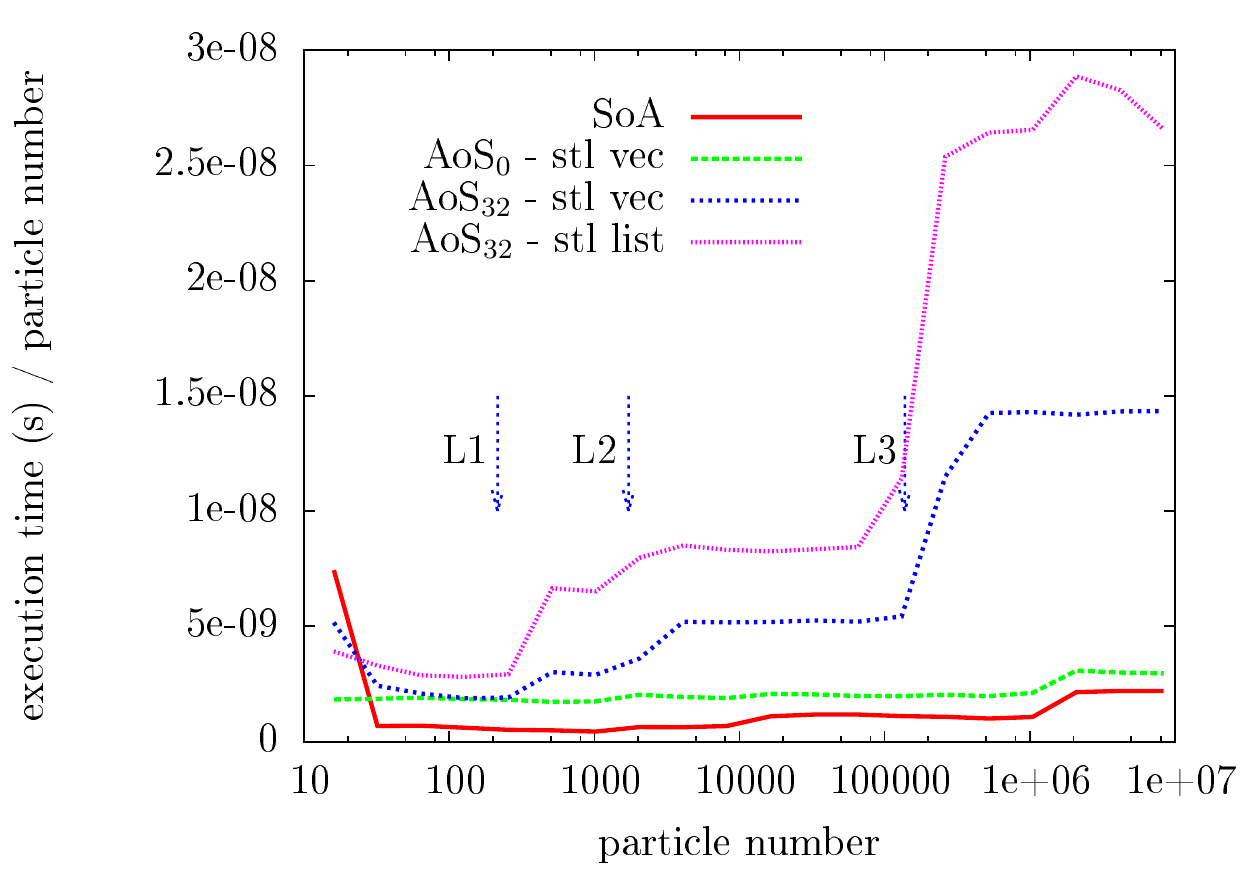}
  \caption{\label{bench} Benchmark comparing the performance of a
    structure of array (SoA) and an array of structure (AoS) on a
    Intel Xeon E5-2680 v3 (Haswell EP). The index SIZE in
    AoS$_\text{SIZE}$ denotes the number of supplementary floats in
    the structure Particle (List.~\ref{lst:particleBench}).}
\end{figure}

\begin{figure}[h]
  \centering
  \includegraphics[width=0.48\textwidth]{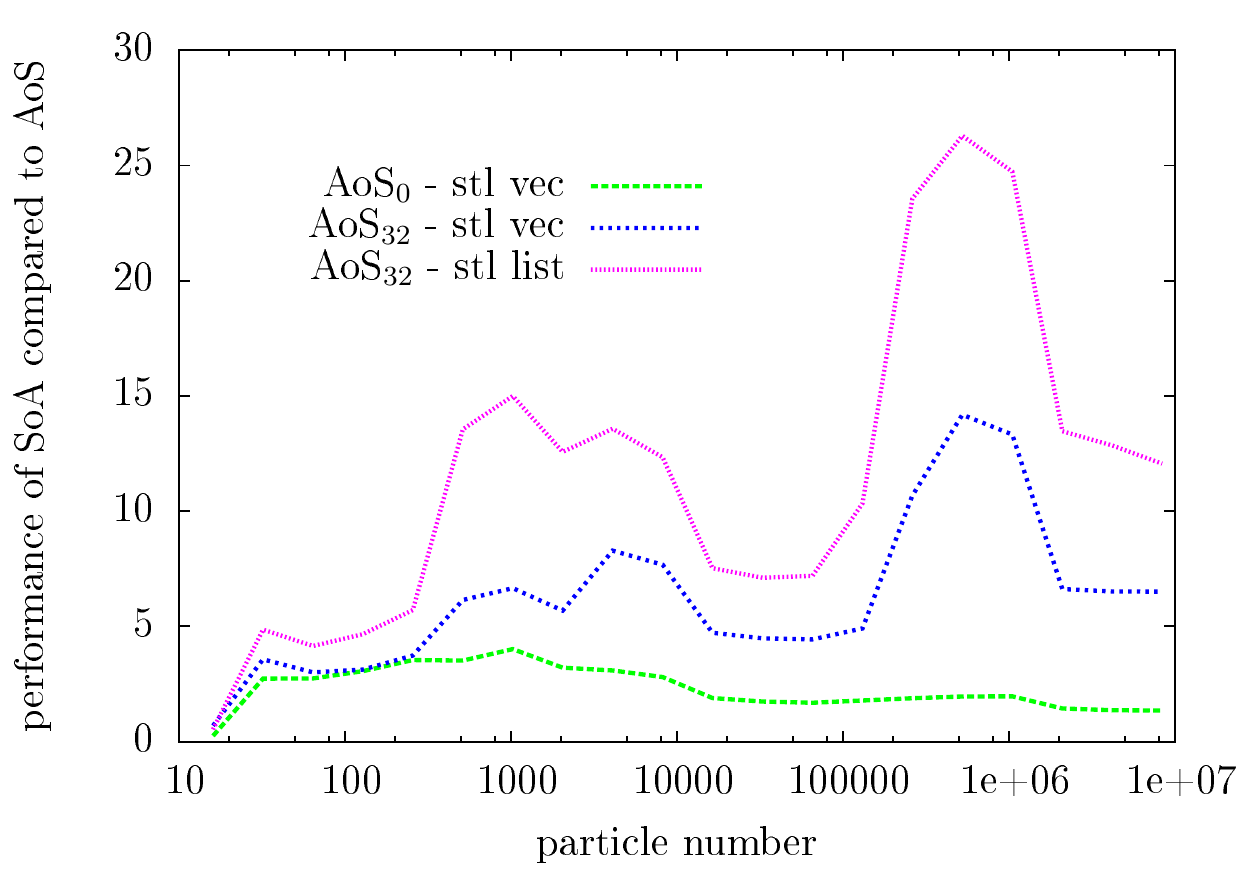}
  \caption{\label{bench_rel} Execution time of an array of structure
    (AoS) relative to that of a structure of array (SoA) on a Intel
    Xeon E5-2680 v3 (Haswell EP). The index SIZE in AoS$_\text{SIZE}$
    denotes the number of supplementary floats in the structure
    Particle (List.~\ref{lst:particleBench}).}
\end{figure}

Figure \ref{bench} compares the normalized execution time for the SoA
and AoS as a function of the particle number. The SoA is much faster
than the AoS. Their relative performance is shown in
Fig.~\ref{bench_rel}. The SoA implementation is up to 25 times faster
than the AoS and one gains at least a factor of two to three by using
a SoA instead of an AoS.

The measured performance depends on the number of particles which is a
consequence of the different cache levels of modern CPUs. Usually they
provide three levels with sizes of 32 kByte (L1), 256 kByte (L2), and
8-40 MByte (L3). The colored arrows in Fig.~\ref{bench} show the cache
limits in terms of a the number of particles of a certain size (in
terms of bytes). One observes that the performance is maximal when the
L1 cache is filled and all particle data still fits into the L2
cache. When the particle data size exceeds the L2 cache, the execution
time slightly increases. An important performance drop happens when
data becomes larger than the L3 level. At that point data has to be
transferred from the main memory that has a significantly lower
bandwidth than the caches.

An important drawback of AoS is that its performance depends on the
size of the particle structure. The more data (properties) this
structure holds, that is to say the bigger it is, the more it fills
the cache that in turn hinders performance. A particle with 32
additional floating point member variables (SIZE=32
List.~\ref{lst:particleBench}) in is much slower than its slim
counterpart. This problem is of course absent for SoA as all arrays
are allocated individually and continuously in memory. Data (particle
properties) that is not used in the execution loop will not be loaded
into the cache.

The execution time of the AoS also depends on the container used to
store the particle objects. A stl vector is significantly faster than
than a stl list. We measure roughly a factor of two. This difference
is due to the additional indirections involved for linked lists (such
as the stl list). On the other hand, a list is faster in removing
particles than a vector as the latter copies successive elements to
keep the data continuous in memory. This drawback can be overcome when
the ordering of particle is not important. In that case, a particle
can be removed by simply overwriting it with the last particle. This
strategy is used by default by SoAx.

\begin{figure}[h]
  \centering
  \includegraphics[width=0.48\textwidth]{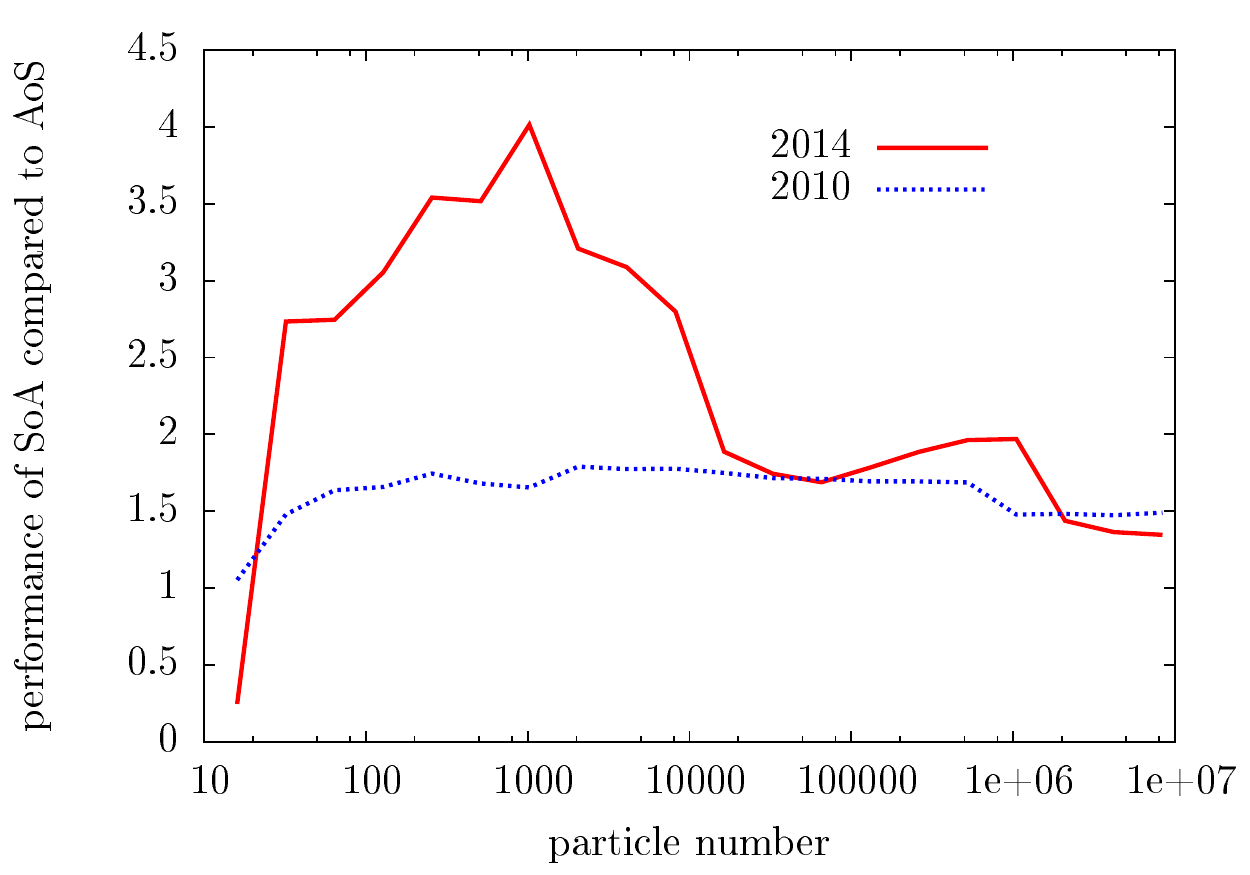}
  \caption{\label{bench_arch} Benchmark comparing SoA and AoS for
    different CPU generations distinguished by the date of their
    commercial launch. 2014: Intel Xeon E5-2680 v3 (Haswell EP); 2010:
    Intel Xeon X5650 (Westmere EP)}
\end{figure}

The performance measured with a given benchmark naturally depends on
the architecture of the CPU. However, it is important to note that the
just discussed relative performance (SoA vs. AoS) will not or only
weakly depend on the clock speed. But other differences, especially
the vectorization units are important as we will show now.  We will
consider two different CPU architectures distinguished by the date of
their commercial release. This sheds light on how the 'SoA vs AoS'
performance ratio changed over time. We compare the SoA performance to
the maximal AoS performance (using the smallest possible particle size
together with a stl vector). In Fig.~\ref{bench_arch} we compare Xeon
CPUs from 2010 and 2014.  For the two CPU generations SoA clearly wins
over AoS. But the modern chip has a higher performance gain. Over only
four years the gain has nearly doubled. The CPU architecture is more
and more constructed in a way that favors the SoA layout.

\begin{figure}[h]
  \centering
  \includegraphics[width=0.48\textwidth]{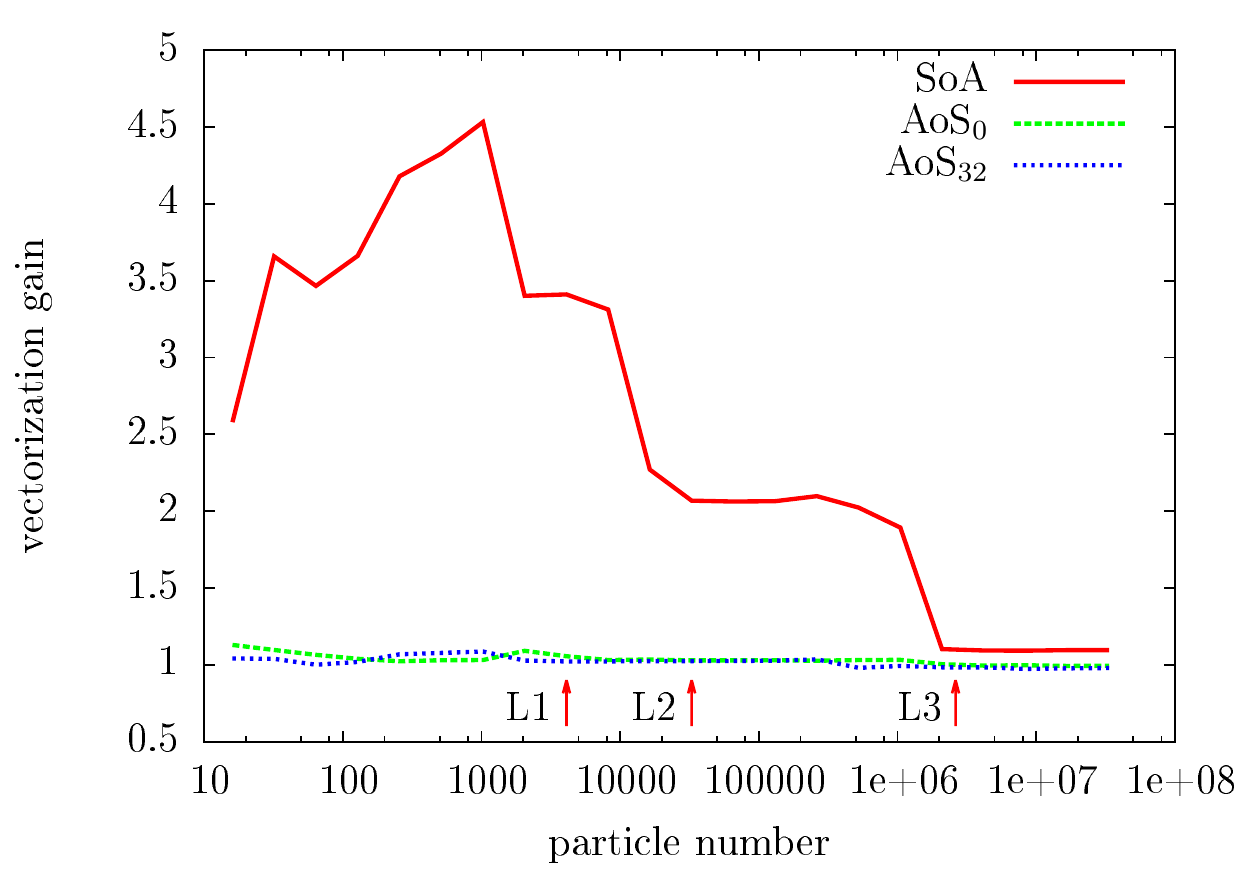}
  \caption{\label{bench_vec} Performance gain due to vectorization for
    a SoA and AoSs on a Intel Xeon E5-2680 v3 (Haswell EP).}
\end{figure}

One architectural component that has changed over the years is the
performance of the vector unit. All today's CPUs possess so-called
single instruction multiple data (SIMD) register and associated
instruction sets. These allow to perform the same instruction (such as
an addition) to many floating-point number at a time (in one cycle)
that can significantly speed up code. In Fig.~\ref{bench_vec} we
compare the performance of SoA and AoS with and without the use of the
vector unit. The vectorization gain of a SoA reaches four to five for
small particle numbers of the order of 100-1000 particles. The
theoretical gain is eight as the used CPU has a 256 bit vector
register containing eight single precision floating point values. At
intermediate particle numbers ($10^3$-$10^6$) the gain is around two
and vanishes for higher particle numbers. The origin of these regimes
can be found in the three cache levels: The gain is maximal if all
data fits into the L2 cache. The second regime corresponds to data
fitting into the L3 cache. However, when the data size exceeds the
latter the vectorization gain vanishes because the data has to be
loaded from the main memory which is too slow to efficiently fill the
vector registers.

Vectorization does not speed up AoS computations. Apparently, the
auto-vectorizer of the compiler does not manage to create a
substantial gain if a AoS is used. This means that a part of the SoA
superiority can be explained by the fact that SoA effectively use the
CPU vector units. 

This also explains the observed differences between the two CPU
architectures. From one CPU generation to the other, the register
width and the set of instruction has been augmented. The old CPU from
2010 has 128 bit vector register with a SSE4.2 instruction set and the
most recent CPU from 2014 has a 256 bit vector register with an AVX2
instruction set. The factor of two between the 128 bit and 256 bit
register explains the differences in Fig.\ref{bench_arch} for
intermediate particle numbers. Of course other features than the
vector unit changed among CPU architectures but it seems that most of
the changes in the 'SoA vs AoS' performance ratio over the years are
due to optimizations of the vector units.

\section{Benchmarks on MICs and GPUs}
\label{sec:bench_GPU_MIC}
Today's supercomputer often use accelerators to speed up
computationally intensive parts of numerical codes. Mainly two
different accelerator types exist:

Intel recently introduced the 'many integrated core' (MIC) concept
with the Xeon Phi co-processor (Knights Corner) that assembles many
computing cores (around 60) on one chip. The used computing cores are
simplified versions of commonly used CPUs so that numerical code
compile without changes on a Xeon Phi.

Nvidia and AMD/ATI developed graphics processing units (GPU) that are
now often used in high performance computing. This architecture uses
hundreds to thousands of very simple computing cores to speed up high
parallel algorithms. For these GPUs the numerical code has to be
especially designed.

The importance of these accelerators for HPC is underlined by the fact
that they are \NOTE{massively employed by the fastest supercomputers
  in the world (according to the TOP 500 list, www.top500.org).}

\subsection{MIC}
\label{sec:bench_MIC}

During the last decade, the performance of supercomputers grew
essentially by increasing the number of (standard) computing cores so
that high performance computing demanded more and more for parallel
numerical algorithms and codes. Intel pushes now further in the
direction of massive parallel programming by introducing
co-processors, called Xeon Phi, with around 60 integrated cores
each. A single core is in general compatible to standard CPUs but
exhibits some architectural differences that are important for the
performance of SoAs and AoSs: A Xeon Phi has no L3 cache but only a 32
kByte L1 and a 512 kByte L2 cache per core. Another aspect is that the
vectorization capacities have been improved by extending the SIMD
registers to 512 bits which means that either 16 single precision
floating point number or 8 double precision number can be processed in
one cycle.

\begin{figure}[h]
  \centering
  \includegraphics[width=0.48\textwidth]{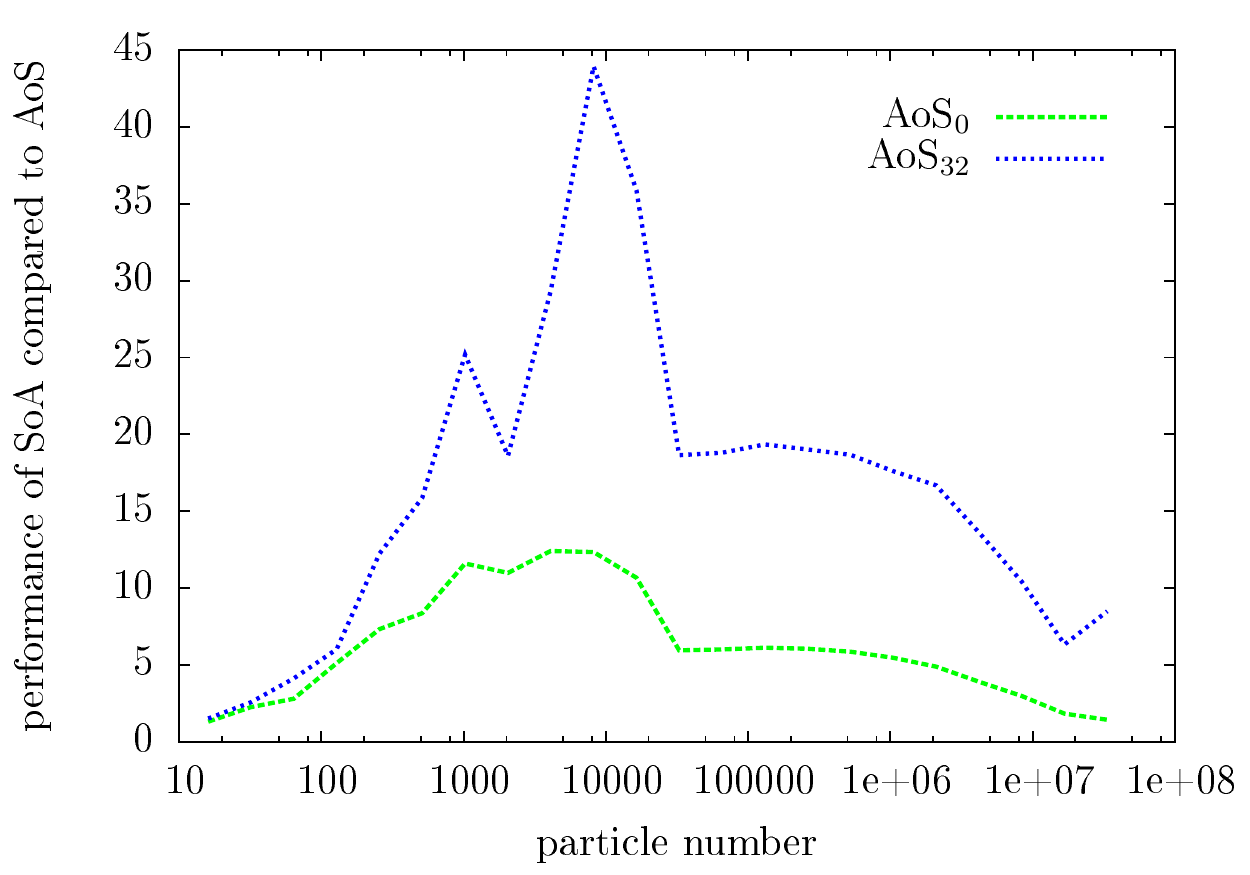}
  \caption{\label{bench_vec_mic} Benchmark comparing the
    performance of AoS to SoA on a Xeon Phi \NOTE{5110P}. }
\end{figure}

These design differences show up in the relative performance of AoS
compared to SoA (as before, we will only study the single-core
performance). Our benchmark \NOTE{(compiled with intel's icc 15)}
shows the the MIC cores favor SoAs over AoS and that even more than
standard CPUs. For small size objects and intermediate particle
numbers the tested SoA is roughly ten times faster than the AoS (see
Fig.~\ref{bench_vec_mic}). If the stored particle has a considerable
size, this difference even varies between twenty and forty.

\begin{figure}[h]
  \centering
  \includegraphics[width=0.48\textwidth]{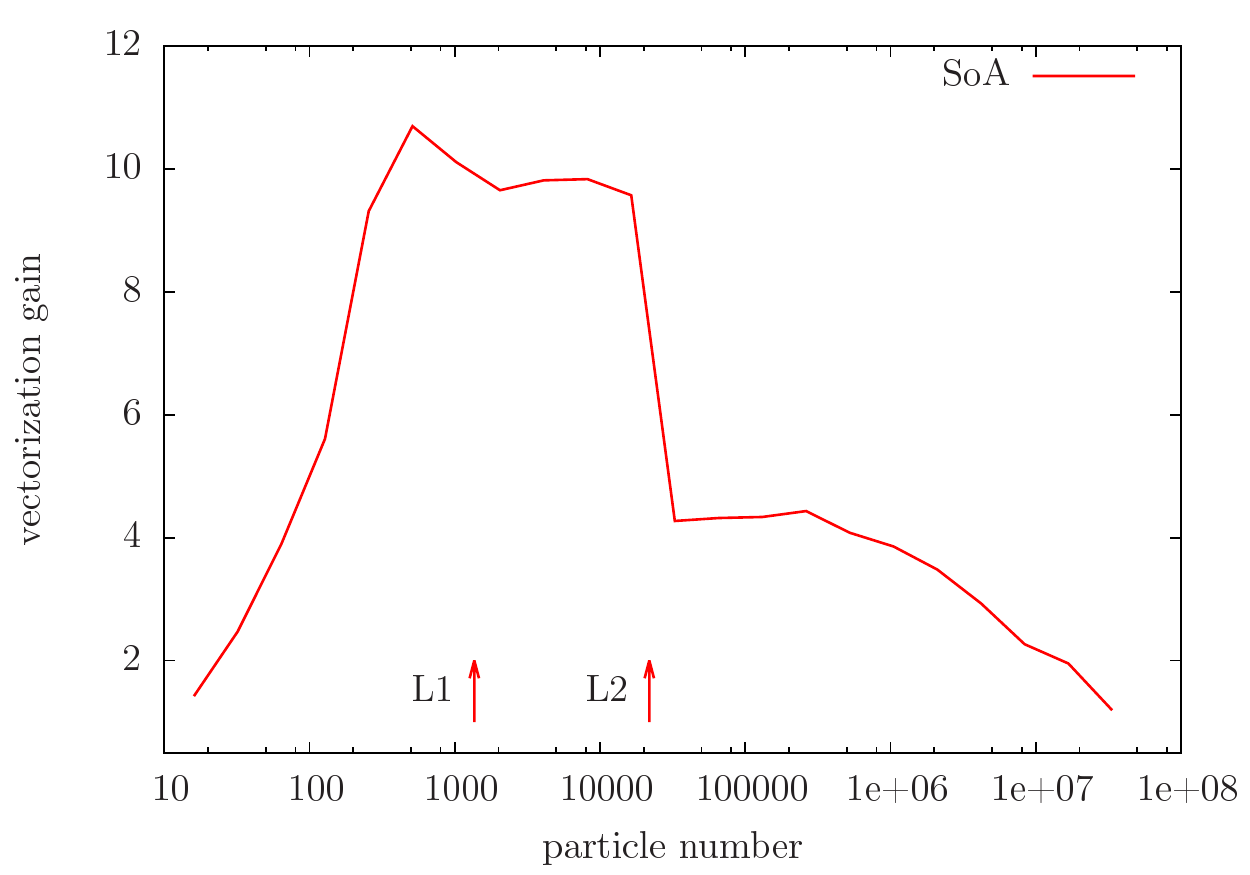}
  \caption{\label{bench_vec_mic_vec} Speed-up by vectorization on a
    Xeon Phi \NOTE{5110P}.}
\end{figure}

The reason is the extended vector performance of the MIC cores. Up to
the point when the L2 cache is filled, vectorization speeds up the
computation by a factor of roughly ten (see
Fig.\ref{bench_vec_mic_vec}) which is below the optimal value of
sixteen but twice the speed-up measured for a standard CPU. Again, the
cache size limits the particle number range for this speed-up.

\subsection{GPU}
\label{sec:bench_GPU}
The architecture behind the General Purpose Graphical Processing Units
(GPU) uses a divide and conquer philosophy, by providing a many-core
device, separated from the CPU, and typically connected to this one
via a PCIe band. Graphic cards are widely used as accelerators in
computer clusters, and power many of the TOP500 fastest supercomputer.

A few thousand of threads can run concurrently on the graphic card,
thus providing the possibility to process many elements at a
time. Furthermore, the architecture, labeled SIMT (for Single
Instruction, Multiple Thread) is somewhat different from the SIMD in
that every single thread has its own register state and can have
independent behaviors from the others, a feature allowing a
thread-based as well as coordinated threads development.

Another important difference from the CPU is the role of the L1
cache. Different caches co-exist, each one belonging to a given
\emph{streaming multiprocessor}, a structure responsible to dispatch
the work among the threads. This cache is mainly used for register
spilling and some stack variables. It does not promote temporal
locality so that repeated operations on the same memory locations will
not necessarily benefit from this cache. The L2 cache, shared among
all streaming multiprocessors, will be used instead. We thus expect
the SoA pattern not to benefit from the L1 cache, but the AoS will in
fact benefit from it : indeed, loading a large structure into memory
allows threads to reuse close memory.

\begin{figure}[h]
\centering
\includegraphics[width=0.48\textwidth]{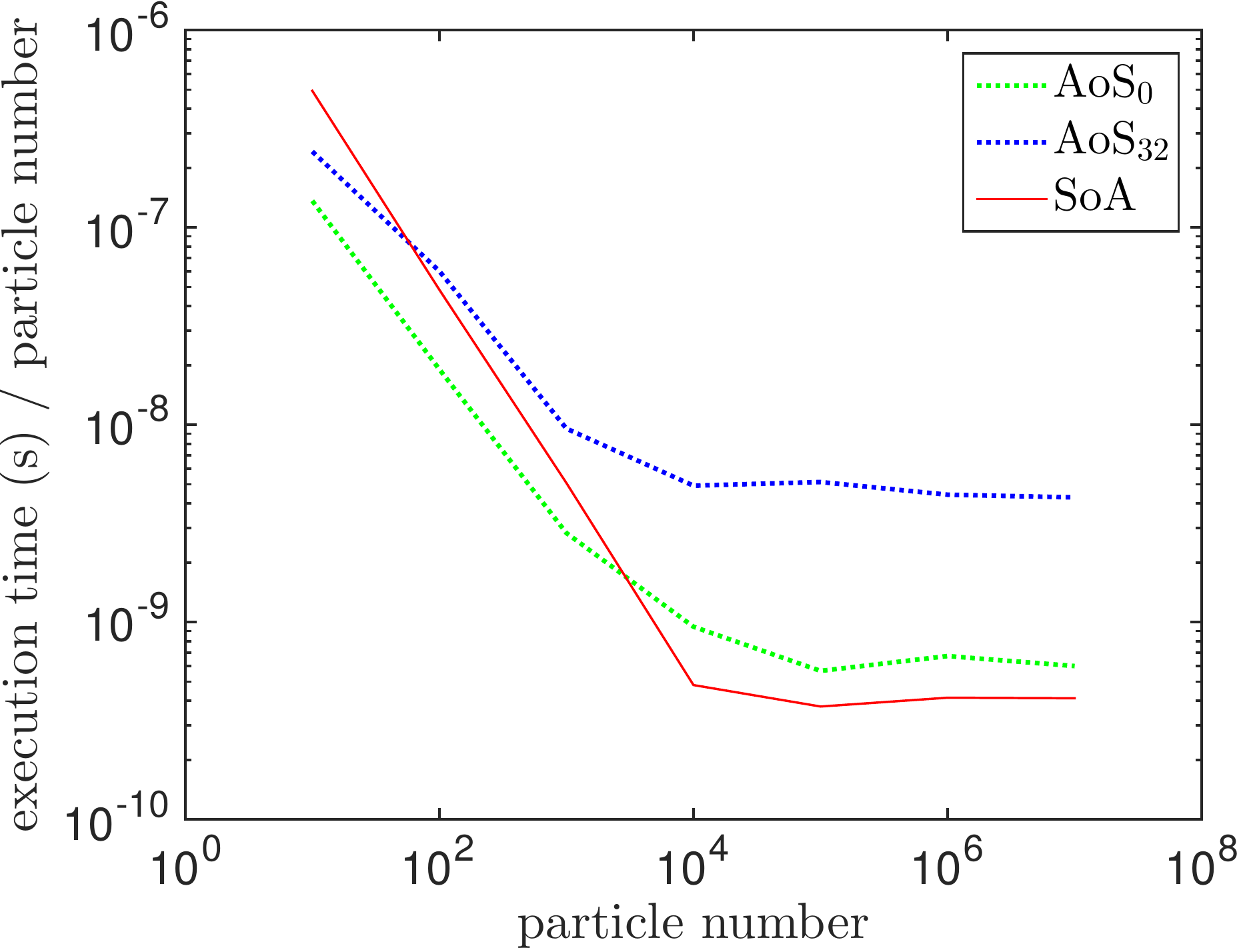}
\caption{\label{fig:bench_gpu_m2050} Benchmark comparing the execution time of AoS vs SoA implementation of the Eulerian update step in single precision.}
\end{figure}

A benchmark similar to those listed above is performed. The graphic
card used is a Nvidia Tesla M2050, a middle-range, widespread
computing device. The card has 448 cores, spread among 14
multiprocessors and the L2 cache size is $\sim 786$ kBytes. In the SOA
algorithm, three functions are launched, one per position and velocity
component, with a number of threads such that each thread has a single
element to process. \NOTE{The version of the used CUDA library is 7.0}. The program is compiled with optimization. Timing
is measured by the Nvidia profiling tool, allowing to isolate the
kernel execution time from the overhead of the function
calls. Execution times normalized by the number of particles are shown
in Fig.~\ref{fig:bench_gpu_m2050}. For large particle numbers, SoA
outperforms AoS solution, by a factor $\sim 2$ for $SIZE=0$ and $~\sim
20$ for $SIZE=32$. As the particle number decreases however, AoS
performs better, with higher crossover for lower $SIZE$. The reason
for this lies in the GPU architecture, as we will now explain.

\begin{figure}[h]
\centering
\includegraphics[width=0.48\textwidth]{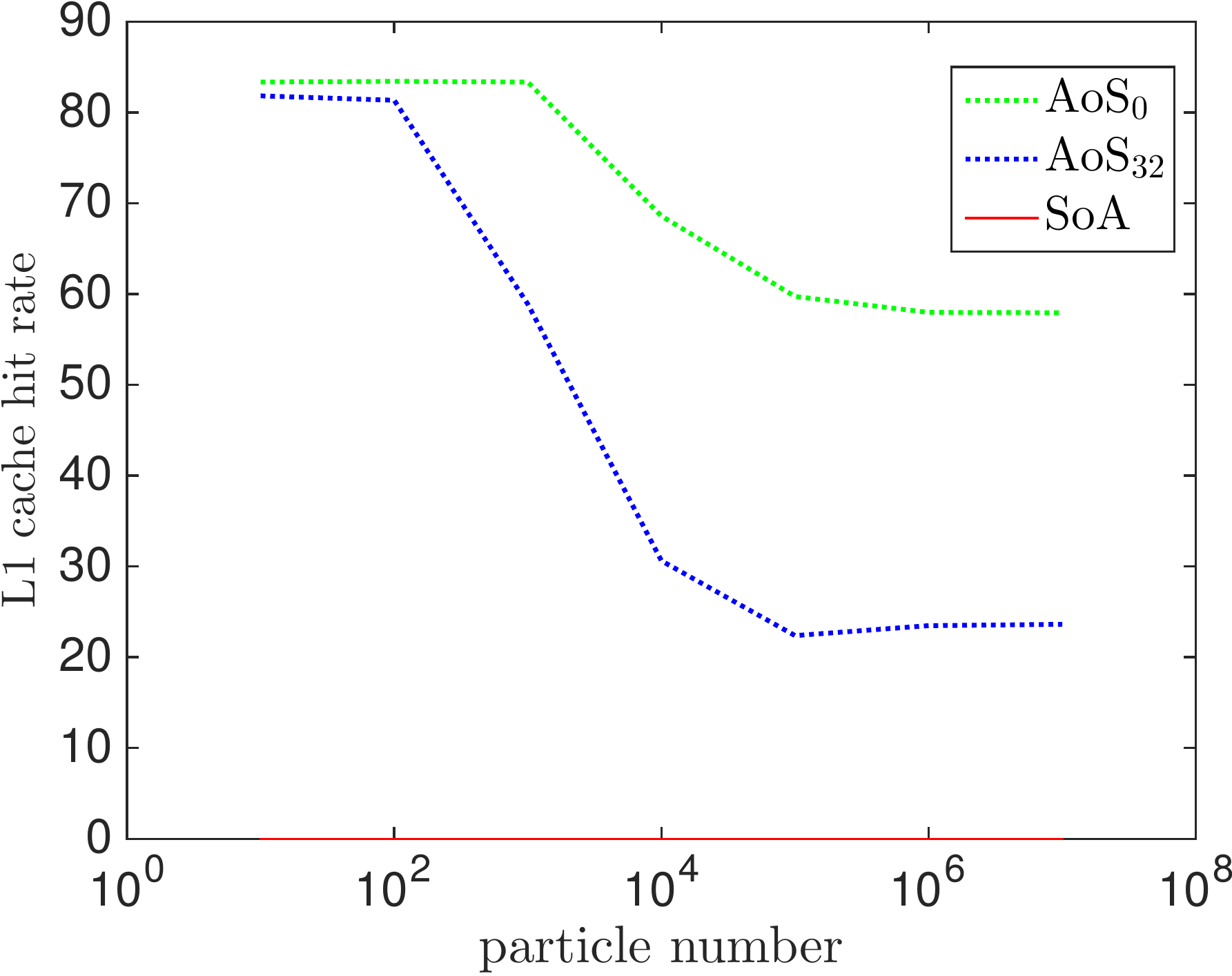}
\caption{\label{fig:metrics_gpu2} Loading efficiency from the main GPU memory. This is the ratio between requested memory and effectively used memory.}
\end{figure}

\begin{figure}[h]
\centering
\includegraphics[width=0.48\textwidth]{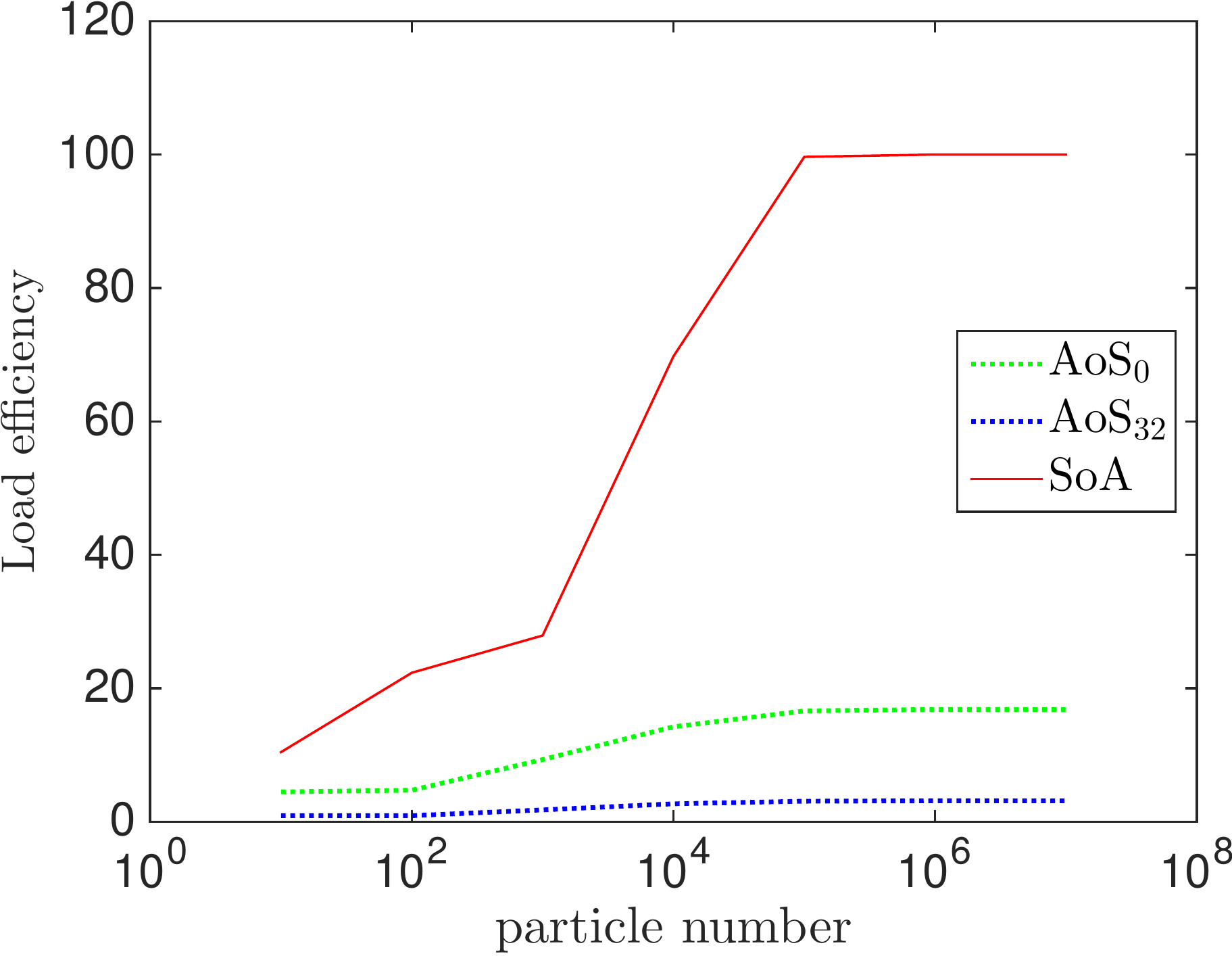}
\caption{\label{fig:metrics_gpu} L1 cache hit rate for global memory load requests, in percents.}
\end{figure}

The major drawback of the AoS approach is the well know effect of
uncoalesced memory access, hence threads fetch unneeded data in the
cache lines. This is particularly damageable in the case of GPU
computing because the major weak point is the latency of memory
access.  Accessing data is done by a single, indivisible group of 32
threads, called a \emph{warp}. Loading a large structure in a thread
memory, only to read a small part of it, degrades badly the memory
access performance up to a factor of 32. The case AoS with $SIZE = 0$
packs 6 values and will then have a memory performance of $1 / 6
\approx 16 \%$ compared to SoA, and the highest values of $SIZE$ will
display a performance down to $1 / 32 \approx 3 \%$. This is shown in
Fig.~\ref{fig:metrics_gpu2}.  As a result, one can clearly see that
the performance per particle saturates for a sufficiently large number
of particles, with SoA pattern outperforming the AoS with $SIZE=32$ by
a factor of 20 and the AoS with $SIZE=0$ by a factor of 2.  For small
particle numbers, performance is hindered by a less effective usage of
memory, additional to the uncoalesced access pattern, as can be seen
in Figure~\ref{fig:metrics_gpu}.

It is also noticeable that the performance of SoA is slightly worse
than AoS for small particle numbers (up to ~1000). This can be
attributed to the fact that when the number of particles is small
enough, the L1 cache and the threads registers are large enough to
keep the whole particles close in memory, hence allowing faster access
to other position and velocity components for successive operations,
while the SoA pattern has to make a request to global memory for every
needed data. Nevertheless, this effect only brings advantage when the
particle number is small. When this number increases, the cache cannot
hold the data anymore and so that the global memory is used and
another long latency fetch has to be performed. The caching advantage
is thus eventually taken over by the poor memory access performance,
and the crossing between SoA and AoS (with $SIZE=0$) occurs around
$~2000$ particles. This corresponds to a full utilization of the L1
cache which is 48 kB, the size of one SoA ($SIZE=0$) particle being
$6*4 = 24$ bytes.

We also performed this simple benchmark on another multi-purpose
graphic card boarded on a desktop computer. For this example, we used
the Nvidia Geforce GT755M, composed of 384 cores on 2 multiprocessors,
with $\sim 262$ kbytes. Timings were $2-3$ times slower, irrespective
of the number of particles and both for AoS and SoA (not shown here),
illustrating the benefit of using a graphic card specifically
dedicated to high performance computing exhibiting more parallelism.

\NOTE{The benchmarks have been performed with relatively old graphic
  cards. However, although the performances are expected to be better
  for both SoA and AoS cases with a more modern graphic card, we do
  not expect the qualitative comparison between these two memory
  layouts to vary. Modern GPUs are capable of exposing more
  parallelism, which we expect to result in an even greater gap
  between the two memory layouts.}



\section{Generic \C++ implementation of a structure of arrays (SoAx)}
\label{sec:soax}
In the introduction we have seen that implementing, maintaining and
using a structure of array can be annoying. We present now an
implementation of a structure of array using modern \C++ (in fact
\C++11), called SoAx (see https://sourceforge.net/projects/soax for
updates and bugfixes), that provides a handy interface, high
flexibility and optimal performance. We use \C++ because is enables
powerful mechanisms to build abstractions without loss of
performance. We discussed in the introduction that adding a property
(such as a charge) to a particle requires the modification of all
member functions (such as PartArr::allocate) that handle the different
arrays. \C++ allows to pass this task to the compiler. Using
\emph{template meta programming}~\cite{abrahams-gurtovoy:2005}, the
needed code can be automatically generated during the compilation. The
result is a class that contains an array for each particle property,
the associated access functions and member functions that allow
efficient handling of all arrays.

\subsection{Using SoAx}
Before discussing details of the implementation let us first show a
short listing presenting some functionality of SoAx. Let us assume
that we want our particles to have an identity, a position, a
velocity, and a mass of types \verb+int+, \verb+double+,
\verb+double+, and \verb+float+, respectively. Let us further assume
that we need three-dimensional coordinates for the position and
velocity. Here is what one could write using SoAx:

\begin{lstlisting}[mathescape=true,caption={\label{lst:usage} Example code showing typical usage of SoAx}]
// Define particle properties through macro
SOAX_ATTRIBUTE(id, 'N');       // identity 
SOAX_ATTRIBUTE(pos, 'P');      // position 
SOAX_ATTRIBUTE(vel, 'V');      // velocity
SOAX_ATTRIBUTE(mass, 'M');     // velocity 

// Specify types and dimension and 
// concatenate attributes using std::tuple
typedef std::tuple<id<int,1>, 
  		   pos<double,3>, 
  		   vel<double,3>,
  		   mass<float,1>> ArrayTypes;

// create SoA for 42 particles
Soax<ArrayTypes> soax(42); 

// access properties of particle 23
soax.id(23) = 0;         // set identity        
soax.pos(23,0) = 3.14;    // set x-coordinate

// operations on all particles ($x = v_y - v_z$)
soax.posArr(0) =soax.velArr(1)-soax.velArr(2);

// allocate memory of 100 particles
soax.resize(100);
\end{lstlisting}

We have payed attention to the fact, that user might want to extract
and treat particles as objects (in the spirit of
\verb+struct Particle+). With SoAx one can write

\begin{lstlisting}[caption={\label{lst:soax::element} Example of using SoAx elements}]
  auto particle = soax.getElement(7);
  particle.id() = 42;
  particle.pos(0) = 3.14;
  soax.push_back(particle);
\end{lstlisting}

The necessary class from which the particle objects are created is
also automatically created by the compiler by means of template meta
programming. This technique will be discussed in the next section.

\subsection{Implementation of SoAx}
SoAx uses \emph{inheritance} in combination with \emph{template
  meta-programming}. The basic idea is to inherit all arrays (particle
properties) into one single structure. The different property types of
the particle are passed to the SoAx class using std::tuple. This is a
component of \C++11 storing heterogeneous data types.

A SoAx attribute consists of an array for storing and member-functions
for accessing data. We have chosen to generate these attribute classes
by macros to avoid repetitive implementations as they have all the
same structure. Macros permit to give custom names to the attributes:
From \verb+SOAX_ATTRIBUTE(pos, 'P');+ the compiler creates a class
with a member-function \verb+pos+ to access individual particles and
\verb+posArr+ to access directly the complete array. The character
\verb+P+ is only a descriptive string that can be used by the user for
other purposes. \verb+pos<double,N>+ is an instantiation of the class
template holding a N-dimensional array of type \verb+double+.

Let us here mention that advanced programming techniques can be used
to provide usage safety. The dimensionality is for example
automatically taken into account for the member function
\verb+pos+. In the case of \verb+pos<double,3>+, \verb+pos(42,0)+
gives the expected access to the first coordinate of particle 42 while
\verb+pos(42)+ yields a compile-time assertion (through 'substitution
failure is not an error' (SFINAE,
\cite{vandevoorde-josuttis:2002})). The behavior is the opposite in
the case of \verb+id<int,1>+, where \verb+id(42)+ is the identity of
particle 42 and \verb+id(42,0)+ results in a compile-time assertion.

Advanced programming techniques also allow to enable the library user
to write automatically optimized code. The line
\verb+soax.posArr(0) = soax.velArr(1)-soax.velArr(2);+ in
List.~\ref{lst:usage} performs an operation on all particles. The
library user does not need to write a custom for-loop for CPUs or a
CUDA kernel for GPUs. For this, SoAx uses a technique called
\emph{expression
  templates}~\cite{veldhuizen:1995,vandevoorde-josuttis:2002} where a
computation such as a sum is encoded in a template. Chained arithmetic
operations are analyzed at compile time and an optimized code without
unnecessary copies is generated by the compiler. This technique is
nowadays used in linear algebra software \cite{aragon:2014}.

\subsubsection{Adding functions}
The user can easily add custom functions to SoAx that he wants to be
applied to all arrays. For this, it is not necessary to touch the code
of the library. The user only has to define a structure containing a
\verb+doIt+ member-function (see List.~\ref{lst:setToValue} for an
example). The first template parameter of this member \verb+doIt+ is a
reference to one of the SoA arrays. Other parameters can be freely
chosen (internally SoAx uses variadic templates). Here is an example
of a function that sets the values of all arrays to a certain value:

\begin{lstlisting}[caption={\label{lst:setToValue} Example of a function to be applied to all SoAx arrays }]
struct SetToValue
{
  template<class T, class Type>
  static void doIt(T& t, Type value) {
    for(int i=0;i<t->size();i++)
    t->operator[](i) = value;
  }
};
\end{lstlisting}
Passing this function to a SoAx object \verb+soax+ as a template
argument,\\ \verb+soax.apply<SOAX::SetToValue>(42);+ applies
\verb+SetToValue::doIt+ to all arrays in \verb+soax+.

This is achieved via recursive templates.  We discuss this programming
technique here as a showcase for the \verb+doIt+ function as it
explains how templates can be used to make the compiler generate code
without loss of performance (see List.~\ref{lst:tupleDo}). In fact,
the SoAx member-function \verb+apply+ calls the member-function
\verb+doIt+ of the class template \verb+TupleDo+ with the particle
attribute tuple (\verb+Tuple+), its size (\verb+N+) and the user
defined template (\verb+DoItClass+ = e.g. \verb+SetToValue+) as
template arguments. The member-function \verb+doIt+ calls recursively
\verb+TupleDo::doIt+ for the attribute tuple but passing a decremented
size. This recursion continues until the passed size is one so that
the compiler chooses the partially specialized case below. Its
\verb+doIt+ member-function calls the \verb+doIt+ function of the user
provided \verb+DoItClass+ that terminates the treatment of the first
entry of the attribute tuple \verb+Tuple+. After that the
\verb+DoItClass::doIt+ is called for the second entry. This process
continues for all attributes. As the code for all calls is generated
at compile time, there is no performance overhead compared to a
hand-written code.

\begin{lstlisting}[caption={\label{lst:tupleDo} Example explaining compile time code creation by recursive templates}]
template<class Tuple, std::size_t N, class DoItClass>
struct TupleDo {
    
  template<class... Args>
  static void doIt(Tuple& t, Args... args)
  {
    TupleDo<Tuple, N-1, DoItClass>::doIt(t,args...);
    DoItClass::doIt(std::get<N-1>(t),args...);
  }
};

template<class Tuple, class DoItClass>
struct TupleDo<Tuple, 1, DoItClass> {
  
  template<class... Args>
  static void doIt(Tuple& t, Args... args)
  {
     DoItClass::doIt(std::get<0>(t),args...);
  }
};
\end{lstlisting}

\subsubsection{GPU implementation}
	
Several restrictions apply when working with GPU processors.  A first
one is the costly data transfer between CPU and GPU: one has to design
a solution in which those transfers are minimized.  Data should reside
mainly on the GPU and be transferred to the main memory only when
needed by the CPU, for example for output to a hard drive. One thus
cannot make use of solutions that would results in dereferenciation by
the CPU of each elements one at a time, but must rely on device
functions that process all data at once on the device. In addition,
when processing multiple vectors with several operations, processing
them all together is faster than successively, an optimization
sometimes referred to as \emph{loop fusion} (see Wikipedia for an
example). These constraints lead us to make again use of expression
templates for device data.

Another constrain comes from the fact that C++-stl vectors are
not designed to work on GPU processors within the CUDA
framework, as far as the version 7.0, and another type of data
storage is then needed.  To allow expression templates to work
with GPUs, we build a custom class, called deviceWrapper,
encompassing a pointer to data living on the device. In
addition, we used the THRUST library \NOTE{\citep{bell2011thrust}, version 1.8.0,} as it provides the best mimic of stl
vectors structure and algorithms to our knowledge. This allows us to
keep trace of the associated device vector to allow efficient
operations to be performed on the data.

When an assignment (of the form
\verb+soax.posArr(0) = soax.velArr(1)-soax.velArr(2);+) is
performed, a kernel is called and passed a copy of the
underlying deviceWrapper object, accessing the data with the
expression template objects. The copy constructor of the
deviceWrapper class then needs to be overloaded in order to
copy \emph{only} the raw device pointer and not all the data
at each call.

\begin{figure}
\centering
\includegraphics[width=0.48\textwidth]{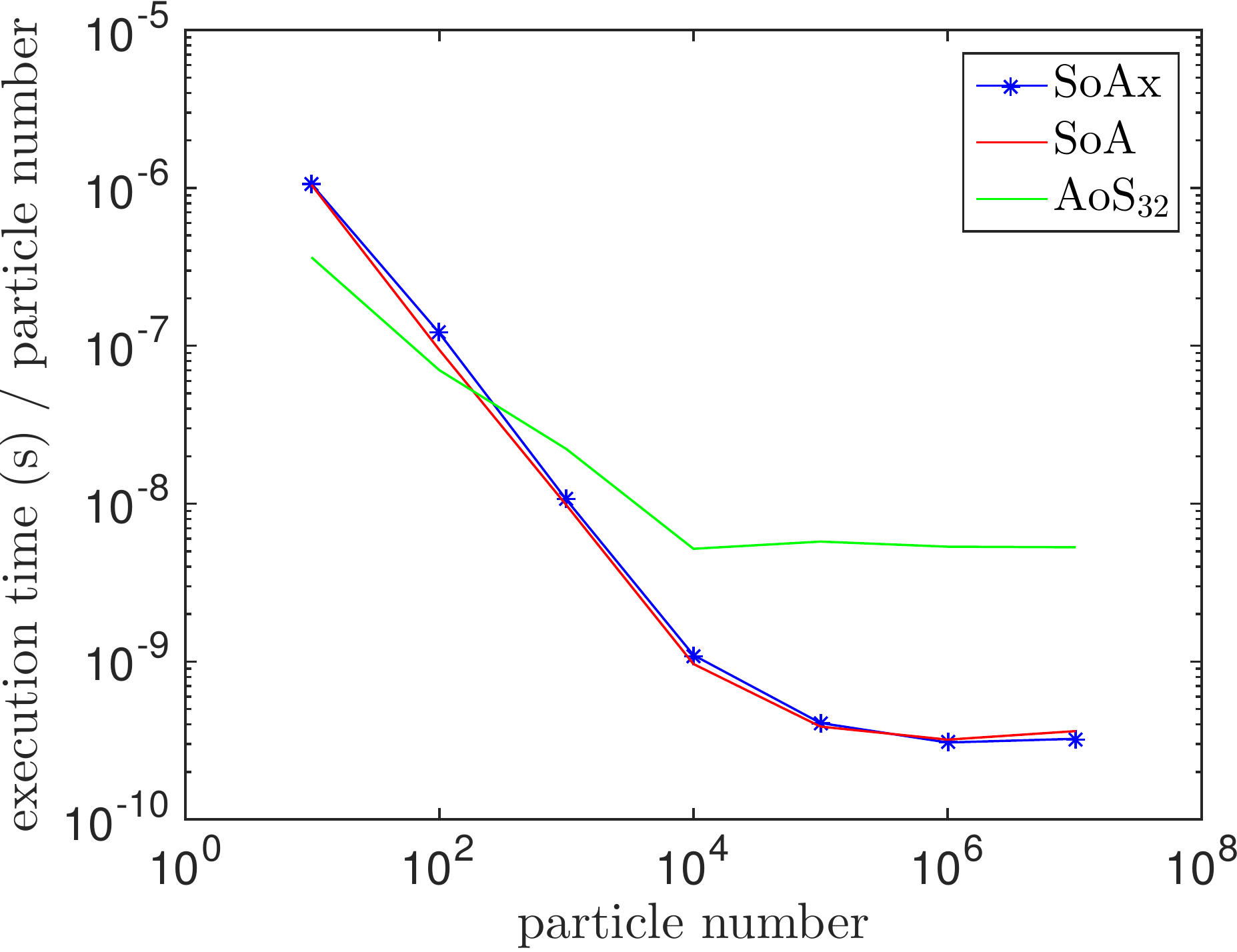}
\caption{\label{fig:benchmark_soax} Benchmark of SoAx library GPU implementation.}
\end{figure}

Fig.~\ref{fig:benchmark_soax} shows a benchmark evaluating the
performance of this implementation for the operation
(\ref{eq:euler}) as a function of the particle number, along
with the SoA and AoS \NOTE{(with $SIZE = 32$)} implementations as
references. The time is measured this time with a std::chrono
rather then with the kernel profiler, allowing to assess the
possible overhead of the SoAx solution. \NOTE{The version of the used CUDA version is 7.0}. 
With this benchmark, we confirm that the performance of SoAx is the same as the SOA
also on GPUs. Indeed, the SoAx GPU implementation comes down
\emph{in fine} to call a kernel on the stored data addressed
through expression templates.

\section{Conclusions}
\label{sec:conclusion}

The goal of the work is two-fold. First, it shows that heterogeneous
data (such as particles) should be implemented in an \emph{array of
  structure} (AoS) fashion rather than in a \emph{structure of array}
(SoA) one if performance is crucial. AoS are generally much faster on
modern CPUs as well as on GPUs. The reason is that AoS better uses
cache and vectorization resources that can speed up typical number
crunching algorithms on particles by more than one order of
magnitude. However, implementing and maintaining AoS can be cumbersome
especially if the the number of numerical types representing a
particle change from one application to another. SoA are in general
more handy and flexible. This consideration leads to the second
contribution of this work showing that modern \C++ programming
techniques permits to combine the advantages of both concepts (SoA and
AoS) to build a generic library that has the performance of SoAs and
the flexibility and handiness of AoS. We demonstrate the benefit of
template meta programming for scientific codes. This technique
delegates code generation to the compiler and allows for highly
readable, maintainable and fast application code. The presented
library SoAx runs on CPUs as well as on GPUs.

\section*{Acknowledgment}
We thank K. Thust for useful discussion and his help using the MIC
co-processor.  We also thank A. Miniussi for fruitful advises
concerning Template Meta Programming. The research leading to these
results has received funding from the European Research Council under
the European Communitys Seventh Framework Program (FP7 / 2007-2013
Grant Agreement no.  240579). Access to supercomputer Jureca and
Juropa3 at the FZ J\"ulich was made available through project
HBO22. Part of the computations were performed on the 'mesocentre de
calcul SIGAMM' in Nice.





\bibliographystyle{elsarticle-num}



\end{document}

y